\documentclass[12pt,twoside]{article}
\usepackage{fleqn,epsfig,espcrc1}

\usepackage{graphicx}
\usepackage[figuresright]{rotating}

\newcommand{\AmS}{{\protect\the\textfont2
  A\kern-.1667em\lower.5ex\hbox{M}\kern-.125emS}}

\title{{\Large{\bf Particle and Nuclear Physics with High Energy Leptons}}}

\author{M W Krasny\address{Balliol College and Nuclear and Astrophysics
      Laboratory, Keble Road, Oxford OX13RH, UK and IN2P3-CNRS,
      Universit\'es Paris VI et VII, 4 pl Jussieu, T33-RdC, 75252 Paris,
      France}} 
\begin{document}
\nonumber

\maketitle

\begin{abstract}
In high centre-of-mass energy lepton-nucleon 
collisions the space-time time resolution
of partonic processes 
can be {\it fine-tuned} within a 
dynamical range which is unattainable in hadronic 
collisions. Replacing nucleons by nuclei of variable 
atomic number enables one to tune the strength of colour
forces. The experimental program of high energy 
electron-nucleon and its extension to electron-nucleus collisions should
thus give an unique opportunity to experimentally 
explore the transition between the soft and hard interactions of small and
extended partonic systems. Such an experimental program,
which can be realized at DESY and/or BNL with relatively modest 
cost, is 
discussed in this talk \footnote{Plenary talk at the {\it PANIC}
confernce, Uppsala, June 1999}.

\end{abstract}

\section{Introduction}

The space-time charge resolution of the leptonic probe
depends upon the following three kinematical variables:
$x_{Bj}$ - the invariant Bjorken variable, $Q^2$ - 
the invariant 4-momentum 
of the exchanged photon ($W$,$Z$ boson) and  $s$ - the 
centre-of-mass energy. 
These variables can be expressed in terms of  the  incoming and  
outgoing lepton  4-momenta $k$ and $k^{'}$ and by the 
4-momentum of the nucleon $p$:

\begin{center}
 
$Q^2 = -q^2 = -(k-k^{'})^2$  \\

$x_{Bj} = Q^2/(2pq)$    \\

$s = (p + k)^2$  \\ 

\end{center}

In the plane perpendicular to the 
collision axis
the constituents of  hadronic matter  carrying electro-weak charges
can be resolved 
within the distances of $l_t \approx 1/Q$.
The corresponding space-time resolution in the 
longitudinal direction is determined by  the value of the 
invariant Bjorken variable - $x_{Bj}$ and by the value of the 
reference frame-dependent Lorenz-factor of the nucleon (nucleus) -
$\gamma$: 
$l_l \approx t \approx 1/(2 \gamma Mx_{Bj})$.
In high energy collisions $l_l$ and $t$
are strongly correlated:
$(t-l_l) \approx 1/s$  simplifying the 
large-distance structure of hadronic matter to  
``frozen configurations".

Strong interactions provide  
natural scales for the resolution of transverse
distances:  $1/\Lambda_{QCD}$ and
the inverse mass of the $\rho$ meson,  $1/ m_{\rho}$. 
Using these yardsticks  
three kinematical regions can be defined:
\begin{itemize}
\item
the photo-production region (PR)
($Q^2 < \alpha _1 \Lambda _{QCD}^2$ ) where the
large-distance structure of hadronic matter is of importance 
and where the   
quasi-real photon interacts  with  the hadronic matter predominantly 
via the  vector-meson component of its wave function. 
\item
the transition region (TR)
($ \alpha _1 \Lambda _{QCD}^2 < Q^2 <  m_{\rho}^2/ \alpha _2$ )
where a direct coupling of the photon ($W$, $Z$ boson) to a
			  charged parton becomes 
			   important 
\item 
the deep inelastic region (DIS)
($Q^2 >  m_{\rho}^2/ \alpha _2$)
where the  direct coupling of the photon to a charged parton dominate
\end{itemize}
 These kinematical regions have rather fuzzy boundaries corresponding 
 to $\alpha _1$ and  $\alpha _2$ values in the range 
 of approximately 0.1-1.0.

The natural scales of  the resolution of the longitudinal distances
in electron-nucleon and electron-nucleus collisions
are: the size of nucleons $R_N$ and size 
of nuclei $R_A$ defined here 
in a Lorenz-invariant way as distances  over which  the valence
quarks of the nucleon (nucleus) are localised.
These values define three kinematical regions:

\begin{itemize}
\item 
 large $x_{Bj}$ region 
($x_{Bj} > 1/(2MR_N) \approx 0.1 $) where the 
the photon (W,Z boson) interaction with the nucleon (nucleus)
is localised 
within the longitudinal distances smaller 
than the nucleon size 
\item
intermediate $x_{Bj}$ region
($1/(2MR_A) \approx 0.01  < x_{B_j} < 1/(2MR_N) \approx 0.1$)
where the  photon interacts coherently 
within the  longitudinal distances  exceeding  the size of the nucleon
\item
small $x_{Bj}$ region 
($x_{Bj} < 1/(2MR_A) \approx 0.01$ )
where the light-cone-coherent interaction of the photon  extends 
over the longitudinal distances exceeding  the size of the nucleus
\end{itemize}

The $(x,Q^2)$ region accessible to the HERA and earlier fixed
target experiments is shown in Fig. 1.
\begin{figure}[!htb]
 \centering
 \epsfig{file=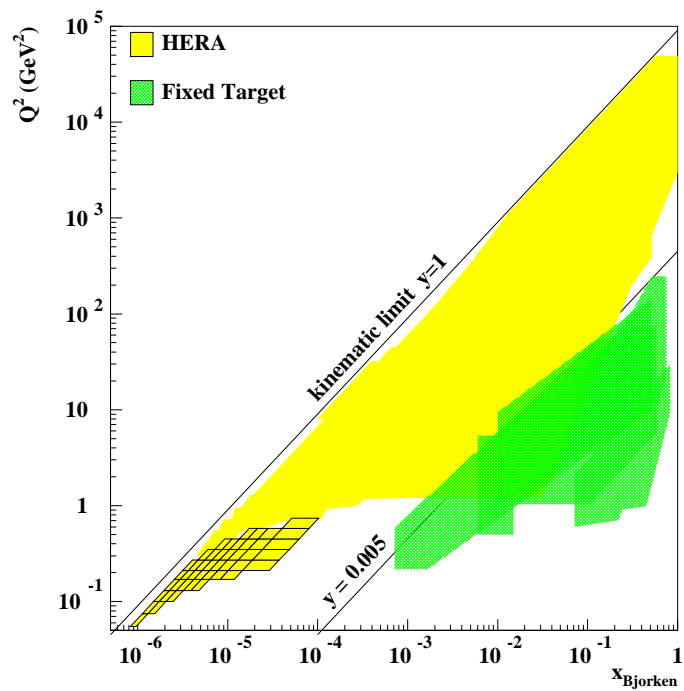,bbllx=86pt,bblly=78pt,bburx=530pt,bbury=670pt,
         width=8cm,angle=0}
\label{fig1}
\caption{The kinematical region covered  by HERA and fixed target
electron-proton scattering experiments.  }
 \end{figure}
The main distinction  between the fixed target and the collider
kinematical domains, resulting from the 
different  centre-of-mass energies,
is that in the latter case the small $x_{Bj}$ region
can be studied in DIS regime where the relevant hadronic 
degrees of motion are quarks and gluons.
In addition,  the HERA collider  experiments
extend the measured DIS region to the $Q^2 > M_W^2$ values
where the Neutral-Current and Charged-Current processes are of comparable 
strength. 

The small $x_{Bj}$, large coherence length, DIS processes can be 
viewed in two equivalent and complementary ways.
In the Bjorken reference frame in which the nucleon (nucleus)
moves with asymptotically large momentum  the photon 
can be considered as colliding  with
delocalised partons, which are described by the light-cone wave function
of the nucleon (nucleus). In this picture nuclei can be considered
as sources of variable ($A^{1/3}$-dependent) 
strength  of the colour fields. The  quark and gluon 
interaction dynamics is expressed in this frame in terms of effective
partonic densities in the nucleon (nucleus).

In the rest frame of the nucleon (nucleus) the small $x_{Bj}$  
DIS collisions can be 
viewed as coherent scattering of various quark-gluon Fock 
components of the virtual photon wave function. In this picture
nuclei can be considered as effective ($A^{1/3}$-dependent) 
filters of various Fock projections of the virtual photon wave function.
The dynamics of quark and gluon 
interactions is expressed in this frame  in terms of the cross
sections for interactions of $q \bar q$, $q \bar qg$ and other Fock
states of the virtual photon with the nucleon (nucleus).

\section{Comments on the present HERA program}

The experimental program  carried out presently at  HERA 
is more exciting  than 
one might think while reading the  
papers and 
listening to the conference presentations of  the 
results approved for public release. 
The relevance of the HERA electron-proton 
scattering program 
and the importance of its extension to
electron-nucleus collisions 
for studies of strong interactions 
can be seen more clearly if one re-formulates
research goals  
from a perspective which is different 
from the HERA orthodoxy.

The change of perspective boils down to using the QED processes 
and the hard
perturbative QCD processes,  
which are theoretically well understood
and cross-measured at $p \bar p$
and $e^{+}e^{-}$ collisions, as tools  to investigate the
{\it QCD-vacuum} and the 
{\it medium} effects in
large (as compared to the confinement scale)
distance propagation of point-like and 
extended colour objects. 
Within such a perspective 
the role of what is considered to be  the ``physics signal"
and what is considered to be 
the ``measurement noise" are reversed. In more practical terms,
instead of getting rid of such non-perturbative effects by 
``re-tuning" of the Monte-Carlo generators or  by absorbing 
them into  
various sets of ``structure functions" 
studies are focused on 
these effects.

The discovery of the rise of the proton $F_2$ structure function
at small $x_{Bj}$  \cite{F2}
provided  experimental evidence
of striking differences in ``large-friction"
propagation of quarks in the "sticky"
QCD medium with respect to a ``frictionless" 
propagation of electric 
charges in ``almost-transparent" QED media.
In the published analyses of the 
structure function data \cite{H1}, \cite{ZEUS}
these effects are absorbed into the 
effective gluon distribution derived in the 
the QCD analyses of the proton structure 
function within the framework of the DGLAP evolution equations. 
The questions if  such a procedure is justified down to low $x_{Bj}$
and if the medium dependent effects can be neglected 
are  open. Changing the medium in which the  photon couples 
to  partons (replacing nucleons by nuclei) provides  
the most straightforward way to
factorise out the medium effects and  to verify the 
applicability of the DGLAP  equations to 
the large density partonic
systems.

One of the most striking medium effects 
observed at HERA is the 
difference in the fragmentation 
of the quarks ejected from the  proton in DIS processes   
with respect to the fragmentation  of quarks
produced in $e^{+}e^{-}$ annihilation processes
\cite{Marchesini}. The Breit-frame spectrum  
of charged hadrons produced in  fragmentation  
of low $x_{Bj}$ quarks is incompatible with  
the corresponding spectrum  measured in 
the  $e^{+}e^{-}$ annihilation.
In addition,  the observed pattern of energy deposition in
the direction of the current low $x_{Bj}$  quarks 
deviates significantly from 
that of the large $x_{Bj}$  quarks  
\cite{Thompson}.   
These observations indicate clearly that the 
medium induced effects  
of quark-energy-loss and quark-multiple-scattering are large
and can be studied quantitatively in a  
future electron-nucleus scattering program.

The large fraction of rapidity gap events observed in  
low $x_{Bj}$ DIS scattering   \cite{rapgap} was unexpected and
remains a mystery if one believes in the validity of the DGLAP 
evolution equations
in this region. The conventional way of analysing these events
within the Regge phenomenology boils down to absorbing
rather than explaining
the source of these events into the pomeron and reggeon
structure functions. 
Studies of rapidity gap events in
electron-nucleus scattering are indispensable in  
understanding  universality  of various  
de-excitations modes of the QCD vacuum.

\section{Electron-nucleus scattering
         at high energies}
	 
High-energy electron-nucleus collisions 
were   discussed at several 
workshops \cite{paris}, \cite{HERA96}, \cite{SEEHEIM}.
Physics highlights of the $eA$ experimental program 
include:

\begin{itemize}
\item studies of the large density partonic
                          systems
			  and  searches for  nonlinear QCD phenomena
\item studies  of partonic structure of the large 
                          distance colour
			  singlet excitations of the  QCD-Vacuum
\item filtering out soft from hard processes  for perturbative QCD studies
\item understanding of colour transparency and colour opacity  			 
\item  studies of  the  space-time structure  of strong interactions
			  using the nucleus as a femto-vertex detector
\item studies of luminous photon-photon scattering 
			  
\end{itemize}
In addition $eA$ collisions provide   
precision measurements of partonic distributions
                          in nuclei in the $x_{Bj}$ range 
			  of $10^{-4}$ - $10^{-1}$
			  and precise 
			  understanding of the  
			  high $E_T$ quark-jets which 
			  can be used as probes of 
			  nuclear medium. They 
			  may turn out to be vital in pinning down
			  the quark-gluon plasma signals 
			  in  the forthcoming analyses of the 
			  RHIC and the LHC-AA data.  
			  
In an  optimal scenario  the atomic numbers of ions
selected for the $eA$ collisions  should  
cover uniformly  the $A^{1/3}$ range  (e.g  $D_2$
       $O_{16}$, $Ca_{40}$, $Sn_{120}$ and $Pb_{207}$)
and the collected luminosities 
should satisfy the condition:               
$L^{eA} \times A \geq  10~ pb^{-1}$/ion. 
The shopping list of measurements 
which can be made if the above criteria are fulfilled
include: 

\begin{itemize}
\item $F_2^A$ and $F_L^A$ structure functions:
                           inclusive and tagged by  the  number
			   of wounded nucleons and evaporation fragments 
\item A-dependence of the gluon distribution 
\item Jet spectra and single particle inclusive
                           spectra in the photo-production, DIS, 
			   and the transition regions
\item A-dependence of the vector meson production  
\item A-dependence of the open charm and beauty production
\item A-dependence of the fraction of rapidity gap events 
\item A-dependence of jet profiles and jet energy loss
\item Fragmentation spectra of 
                           tagged low-$x$ partons in nuclear medium, also in correlation 
			   with the observed  number
			   of wounded nucleons and evaporation fragments
\item Particle multiplicities and particle correlations 
                           in $\gamma ^{*} A$ and $\gamma A$ scattering
			  
\item Bose-Einstein correlations and their A-dependence

\end{itemize}
 
 \begin{figure}[!htb]
 \centering
 \epsfig{file=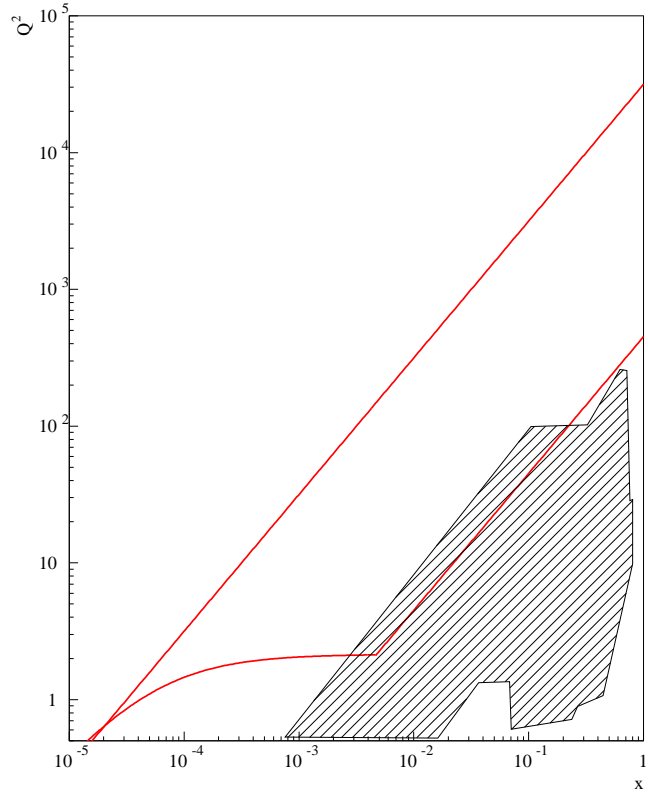,bbllx=86pt,bblly=78pt,bburx=530pt,bbury=670pt,
         width=8cm,angle=0}
\label{bla}
\caption{The kinematical region covered which can be covered in 
electron-nucleus collisions at HERA. The hatched
area corresponds to the region covered in the fixed
target $eA$ experiments.}
 
\end{figure}
 
Electron-nucleus collisions can be realized  at DESY and/or at BNL.
Preliminary machine studies 
presented by J. Maidement at the
``Physics with HERA as an $eA$ collider" workshop  show that with
       relatively modest investment (ion source, RFQ)  isoscalar nuclei   
       of the energies of $410 \times A$  GeV  can be accelerated and stored 
       at  HERA   
       providing luminosities:  
\begin{itemize}
\item 
$L^{HERA}_{eA} \times A = (0.5-1.0) \times L_{ep}^{HERA}
                     \approx 100~ pb^{-1}$/year 
                     for electron - deuteron collisions
\item
$L^{HERA}_{eA} \times A \approx  1/6 \times L_{ep}^{HERA} 
                    \approx 20~ pb^{-1}$/year 
                    for electron - oxygen collisions
\end{itemize}		    
In order to 
collide electrons with heavier nuclei a purpose-built heavy ion preinjector 
system is needed.  For electron - lead  collisions the
expected luminosity is:
\begin{itemize}
\item 		    
$L^{HERA}_{eA} \times A \approx  (1/20 -1/50)\times L_{ep}^{HERA} 
                   \approx 3-7~ pb^{-1}$/year. 
\end{itemize}           
At BNL, preliminary machine studies \cite{Peggs} show that
       it is feasible to collide heavy ions with electrons 
       by building a purpose-designed
       room-temperature electron(positron) ring in the RHIC tunnel.
       As an example, the expected          
       luminosity for eAu  collisions is estimated to be:
\begin{itemize}
\item          
       $L^{BNL}_{eA} \times A = 3.7 \times A \times 10^{29} cm^{-2}s^{-1}
                   \approx  L_{ep}^{HERA} $
\end{itemize}
                     for collisions of  		    
                     10 GeV electrons with 
		     $100 \times A$ GeV ions.

The above  luminosities indicate that the electron-nucleus program 
requires at least 2-4 years of $eA$ collisions  at HERA.
It is worthwhile  mentioning  that 
the simultaneous storage of two or three
       types of isoscalar nuclei allows one to drastically reduce  the 
       systematic uncertainties  of  
        A dependent ratios of various observables  
        to $\approx$ 1 \% level. 
	Such a running mode
	is considered possible by the machine experts at HERA.

What could be the time-schedule for such a  program? If 
the BNL and/or DESY $eA$ projects are approved by the year 2002,
the first electron-ion collisions can be observed in 2005. 
For HERA, the eA collisions would follow 
the ``high lumi" program 
which is expected to start in 2001. 

Two possible  detector scenarios are being considered.
The first consists of building a dedicated detector for $eA$ 
collisions. An alternative, cost-effective solution is to use
the existing detector(s) upgraded for the $eA$ runs.

 \begin{figure}[!htb]
 \centering
 \epsfig{file=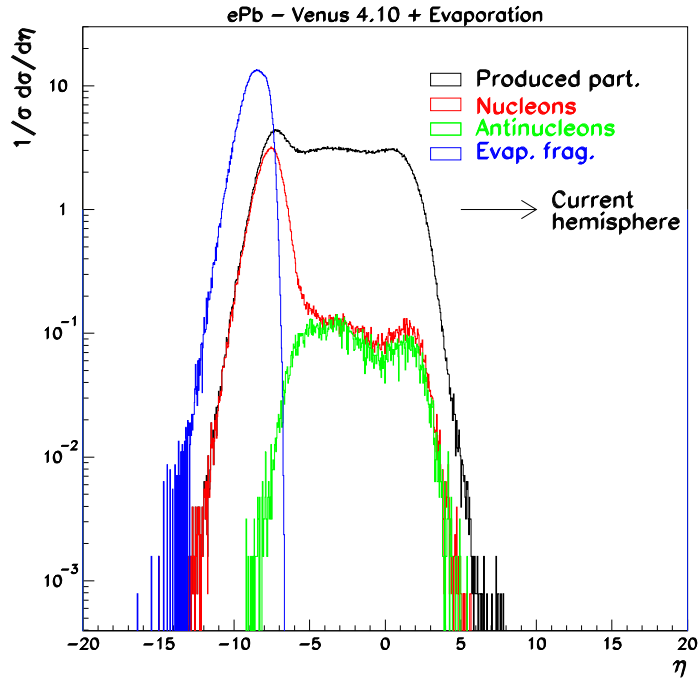,bbllx=86pt,bblly=78pt,bburx=530pt,bbury=670pt,
         width=8cm,angle=0}
\label{eta}
\caption{Pseudorapidity distribution of particles produced in 
electron-lead DIS scattering - Monte-Carlo studies.}  
\end{figure}

The $(x,Q^2)$ coverage of the existing HERA detectors 
for collisions of 27.6 GeV electrons with 
ions accelerated to the energy of 410 Gev/nucleon is shown
in Fig. 2. The $Q^2$ region which can be
measured by these detectors 
is limited by the angular coverage of the central detectors. 
Note, that both HERA detectors \cite{detectors} provide measurements 
of the energies and angles of electrons and photons
emitted collinearly with the incoming electron. 
Consequently, the 
photoproduction events can be  tagged and  radiative processes
can be experimentally controlled within the present detector
set-up at HERA.

In Fig. 3 the pseudorapidity spectrum of particles
produced in collisions of electrons with lead nuclei
is shown. The present HERA detectors
cover the pseudorapidity range down to $\eta \approx -3.5$.
Below this pseudorapidity value the produced 
particles (mostly nucleons and nuclear fragments)
remain undetected. 
This region is particularly interesting
both for $ep$ and $eA$ collisions and the 
upgrade of the existing detectors 
to detect these particles is 
indispensable for the future experimental program.

\section{Conclusions}

Over the last 20 years we have witnessed
                          impressive experimental and theoretical
			  progress in  understanding the 
			   point-like structure
			  of hadrons  and in 
			  understanding short-distance   processes
                    	  involving quarks and gluons.
                          Understanding the role of quarks and gluons 
                          in large-distance, soft
			  processes remains  
			  one of the most important
			  challenges for  the coming decade. 
			  Electro-weak probes of  tunable space-time
                           resolution applied to hadronic media of 
			   tunable  colour force strength
			   provide an  ``adiabatic approach-path"
			  to large-distance processes.
                           The experimental program of high energy
                          electron-nucleus scattering, which can 
			  address these challenges is  exciting, 
			  feasible  
			  and cost-effective.
                          It can be realized with relatively  modest 
                          modifications  of the existing accelerators 
			  at  DESY and/or  BNL,  
			  but needs a joint-effort of the particle 
                          and nuclear physics communities.

\section*{Acknowledgments}

I would like to thank all colleagues from the 
Department of Physics of the Oxford University
and from Balliol College 
who made my sabbatical stay   
so ejoyable. I am indebted particularly to R. Devenish
for his hospitality.

\end{document}